\documentclass[portrait,11pt]{article}

\usepackage[paper=a4paper,dvips,top=2.2cm,left=2.2cm,right=2.2cm,bottom=2.6cm]{geometry}
\usepackage{amsfonts,amsmath,amsthm,amssymb,enumerate,epsfig,graphicx}
\usepackage[svgnames]{xcolor}
\usepackage{fix-cm}
\usepackage{sectsty}
\usepackage{fancyhdr}
\usepackage{lastpage}
\usepackage{graphicx}
\usepackage{url}
\usepackage{paralist}
\usepackage{multicol}
\usepackage{color}  %%% Needs to make colour of some content of your text
\usepackage{float}  %%% Fixes the position of the figure
\usepackage{hyperref}   %%% Needs to make hyper reference of any references or citations
\hypersetup{colorlinks=true,linkcolor=beamer@mediumblue, citecolor=beamer@new}
\definecolor{beamer@blue}{RGB}{0,0,255}
\definecolor{beamer@mediumblue}{RGB}{0,0,190}
\definecolor{beamer@midnightblue}{RGB}{25,25,112}
\definecolor{beamer@navy}{RGB}{0,0,128}
\definecolor{beamer@darkblue}{RGB}{0,0,139}
\definecolor{beamer@purple}{RGB}{128,0,128}
\definecolor{beamer@levander}{RGB}{100.,149.,237.}
\definecolor{beamer@green}{RGB}{0,128,0}
\definecolor{beamer@darkgreen}{RGB}{0,150,0}
\definecolor{beamer@olive}{RGB}{128,128,0}
\definecolor{beamer@darkolivegreen}{RGB}{85,107,47}
\definecolor{beamer@gray}{RGB}{190,190,190}
\definecolor{beamer@ivry}{RGB}{220,220,220}%{238,232,205}
\definecolor{beamer@new}{RGB}{40,120,50}
\definecolor{shadecolor}{RGB}{220,220,220}
\definecolor{beamer@darkslategray}{RGB}{47,79,79}
\definecolor{beamer@chocolate}{RGB}{210,105,30}
\definecolor{beamer@orangered}{RGB}{255,69,0}
\definecolor{beamer@maroon}{RGB}{128,0,0}
\definecolor{beamer@white}{RGB}{234,242,243}
\definecolor{beamer@silver}{RGB}{0.5,0.45,0.37}

\begin{document}
			
%%%%%%%%%%%%%%%%%%%%%%%%%%%%%%%%%%%%%%%%%%%%%%%%%%%%%%%%%%%%%%%%%%%%%%%%%%%%%%%%%%%%%%%%%%%%%%%%%%%%
%  Title
%%%%%%%%%%%%%%%%%%%%%%%%%%%%%%%%%%%%%%%%%%%%%%%%%%%%%%%%%%%%%%%%%%%%%%%%%%%%%%%%%%%%%%%%%%%%%%%%%%%%

\title{\textbf{$q$-deformed noncommutative cat states and their nonclassical properties}}
\author{\textbf{Sanjib Dey}}
\date{\small{Centre de Recherches Math{\'e}matiques (CRM), Universit{\'e} de Montr{\'e}al \\ Montr{\'e}al - H3C 3J7, Qu{\'e}bec, Canada \\ $\&$ \\ Department of Mathematics and Statistics, Concordia University \\ Montr{\'e}al - H3G 1M8, Qu{\'e}bec, Canada \\ E-mail: dey@crm.umontreal.ca}}
\maketitle
    	
%%%%%%%%%%%%%%%%%%%%%%%%%%%%%%%%%%%%%%%%%%%%%%%%%%%%%%%%%%%%%%%%%%%%%%%%%%%%%%%%%%%%%%%%%%%%%%%%%%%%
%  Abstract
%%%%%%%%%%%%%%%%%%%%%%%%%%%%%%%%%%%%%%%%%%%%%%%%%%%%%%%%%%%%%%%%%%%%%%%%%%%%%%%%%%%%%%%%%%%%%%%%%%%%

\thispagestyle{fancy}

\begin{abstract}
We study several classical-like properties of $q$-deformed nonlinear coherent states as well as nonclassical behaviours of $q$-deformed version of the Schr{\"o}dinger cat states in noncommutative space. Coherent states in $q$-deformed space are found to be minimum uncertainty states together with the squeezed photon distributions unlike the ordinary systems, where the photon distributions are always Poissonian. Several advantages of utilising cat states in noncommutative space over the standard quantum mechanical spaces have been reported here. For instance, the $q$-deformed parameter has been utilised to improve the squeezing of the quadrature beyond the ordinary case. Most importantly, the parameter provides an extra degree of freedom by which we achieve both quadrature squeezed and number squeezed cat states at the same time in a single system, which is impossible to achieve from ordinary cat states.  
\end{abstract}
	 
%%%%%%%%%%%%%%%%%%%%%%%%%%%%%%%%%%%%%%%%%%%%%%%%%%%%%%%%%%%%%%%%%%%%%%%%%%%%%%%%%%%%%%%%%%%%%%%%%%%%
%  Introduction
%%%%%%%%%%%%%%%%%%%%%%%%%%%%%%%%%%%%%%%%%%%%%%%%%%%%%%%%%%%%%%%%%%%%%%%%%%%%%%%%%%%%%%%%%%%%%%%%%%%%

\begin{section}{Introduction}
Coherent states are interesting superposition of infinitely many quantum states exhibiting classical-like features. It was Schr{\"o}dinger, who first discovered them in 1926 \cite{Schrodinger} in form of nonspreading wave packets of harmonic oscillator. They were derived later in 1963 in a more systematic and formal way as eigenstates of non-Hermitian annihilation operators by Glauber \cite{Glauber}, who coined the term``coherent states" for the first time in the literature. Since then several types and generalizations of coherent states have been proposed and their properties have been analysed; see, for instance, \cite{Dodonov,Ali_Antoine_Gazeau} for reviews on the developments. Coherent states corresponding to Lie groups have found numerous applications in quantum optics \cite{Perelomov,Onofri}, whereas nonlinear coherent states \cite{Filho_Vogel} and $f$-coherent states \cite{Manko_Marmo_Sudarshan_Zaccaria} corresponding to nonlinear algebras have also been very useful in this context. Recently the author, with his collaborators has studied few classical-like properties of generalised coherent states and discussed their superiorities on noncommutative \cite{Dey_Fring_squeezed} and $q$-deformed \cite{Dey_Fring_Gouba_Castro} spaces rather than the usual space. Nevertheless, coherent states are undoubtedly fascinating states of light in which quantum noise is reduced to as little as the noise produced by a vacuum state.

On the other hand, in squeezed states of light, the fluctuation of electric fields at certain phases are minimised to even less than that of a vacuum state \cite{Walls,Loudon_Knight}, which means when we turn on squeezed light we observe less amount of noise than with no light at all. This paradoxical behaviour is a direct consequence of the quantum nature of light, which cannot be explained within the classical framework. However, this amount of success has been achieved after resolving many paradoxes such as the EPR paradox, which was proposed by Einstein, Podolsky and Rosen in 1935 \cite{Einstein_Podolsky_Rosen}. Today quantum entanglement has been observed in many optical \cite{Aspect_Grangier_Roger,Kwiat} and ionic systems \cite{Sackett,Julsgaard} and is recognised as a resource of quantum information processing. 

At about the same time as the EPR discussion, Schr{\"o}dinger proposed his famous cat paradox \cite{Schrodinger_cat}, whose concept was to extend the Copenhagen interpretation of quantum mechanics to macroscopically distinguishable objects. From the quantum optical point of view, coherent states are identified to macroscopic objects and therefore the idea has been extended to superpose two coherent states of the same amplitude but with different phases, which are known as Schr{\"o}dinger's cat states. Cat states have been applied in various branches of physics, in particular in the field of quantum optics as they provide many nonclassical features of light \cite{Gerry_Knight} such as quadrature and amplitude (number) squeezing, photon bunching and antibunching etc. Moreover, they have been attempted to identify as basic states for logical qubit to generate quantum gates \cite{Gilchrist,Ourjoumtsev,Gao}. 

In the present work we study Schor{\"o}dinger cat states in $q$-deformed noncommutative space and investigate their nonclassical properties. We choose here a $q$-deformed algebra which has been shown to be related to the noncommutative space-time structures leading to the existence of minimal lengths and minimal momenta as a result of the generalised uncertainty relation \cite{Bagchi_Fring,Fring_Gouba_Scholtz,Fring_Gouba_Bagchi,Dey_Fring_Gouba}. We identify several advantages of utilising these kind of noncommutative spaces rather than the usual quantum mechanical systems. A striking feature is the existence of both quadrature and number squeezing for the even cat states in noncommutative systems presented here, which is however not achieved in the case of ordinary systems. There are very few authors who studied nonlinear systems in this context, for instance one may look at \cite{Mancini,Wu_Yang}. However, most of the investigations were carried out based on the verification of the nonclassical properties of noncommutative systems similar to ordinary systems \cite{Roy,Sharma_Sharma}. The advantages of choosing noncommutative systems over the usual systems have not been reported notably before.

\lhead{}
\chead{}
\rhead{$q$-deformed noncommutative cat states}

\end{section}

%%%%%%%%%%%%%%%%%%%%%%%%%%%%%%%%%%%%%%%%%%%%%%%%%%%%%%%%%%%%%%%%%%%%%%%%%%%%%%%%%%%%%%%%%%%%%%%%%%%%
% Section 2
%%%%%%%%%%%%%%%%%%%%%%%%%%%%%%%%%%%%%%%%%%%%%%%%%%%%%%%%%%%%%%%%%%%%%%%%%%%%%%%%%%%%%%%%%%%%%%%%%%%%

\begin{section}{$q$-deformed nonlinear coherent states}\label{sec2}
Let us start our discussion by considering a set of generalised ladder operators $A^\dagger$ and $A$ in terms of the bosonic creation and annihilation operators $a^\dagger$ and $a$
\begin{eqnarray}\label{genladder}
A^\dagger &=& a^\dagger f(n)=f(n-1)a^\dagger \\
A &=& f(n)a=a f(n-1),\notag
\end{eqnarray}
where $f(n)$ is an operator-valued function of the Hermitian number operator $n=a^\dagger a$. The operators $A$ and $A^\dagger$ therefore obey the following nonlinear commutator algebras
\begin{equation}\label{defcommutator}
\big[A,A^\dagger\big]=(n+1)f^2(n)-nf^2(n-1), \quad \big[n,A\big]=-A \quad \text{and} \quad \big[n,A^\dagger\big]=A^\dagger,
\end{equation} 
where the nonlinearity arises from $f(n)$. Clearly, with the choice of $f(n)=1$, the deformed algebra (\ref{defcommutator}) reduces to the Heisenberg algebra
\begin{equation}
\big[a,a^\dagger\big]=1, \quad \big[n,a\big]=-a \quad \text{and} \quad \big[n,a^\dagger\big]=a^\dagger.
\end{equation}
In analogy to the Glauber states \cite{Glauber}, the nonlinear coherent states \cite{Filho_Vogel,Manko_Marmo_Sudarshan_Zaccaria,Sivakumar} are therefore defined as the right eigenvector of the generalised annihilation operator $A$:
\begin{equation}\label{defeigen}
A\big\vert\alpha,f\big\rangle=\alpha\big\vert\alpha,f\big\rangle,
\end{equation}
where $\alpha$ is a complex eigenvalue, which is however allowed as $A$ is non-Hermitian. Solving the eigenvalue equation (\ref{defeigen}) one then obtains an explicit expression of coherent state in number state representation  
\begin{equation}\label{noncoherent}
\big\vert\alpha,f\big\rangle=\frac{1}{\mathcal{N}(\alpha,f)}\displaystyle\sum_{n=0}^{\infty}\frac{\alpha^n}{\sqrt{n!}~h(n)}\vert n\rangle, \qquad \alpha\in\mathbb{C},
\end{equation}
where
\begin{equation}\notag
h(n)=\left\{ \begin{array}{lcl}
1 & \mbox{if}
& n=0 \\ \displaystyle\prod_{k=0}^{n-1} f(k) & \mbox{if} & n>0~.\end{array}\right.
\end{equation} 
It is possible to define another set of ladder operators $B$ and $B^\dagger$  \cite{Roy_Roy}, which are canonically conjugate to $A$ and $A^\dagger$
\begin{equation}
B^\dagger=a^\dagger\frac{1}{f(n)} \qquad \text{and} \qquad B=\frac{1}{f(n)}a,
\end{equation}
so that one can easily check $[A,B^\dagger]=[B,A^\dagger]=1$, which allows one to write the displacement operator
\begin{equation}
D\big(\alpha,f\big)=e^{\alpha B^\dagger-\alpha^\ast A},
\end{equation}
and construct nonlinear coherent states through
\begin{equation}
\big\vert\alpha,f\big\rangle=D\big(\alpha,f\big)\vert 0\rangle~.
\end{equation}
The outcome coincides exactly with (\ref{noncoherent}). The normalisation constant can be computed from the requirement $\big\langle \alpha,f\big\vert\alpha,f\big\rangle=1$, so that
\begin{equation}
\mathcal{N}^2(\alpha,f)=\displaystyle\sum_{n=0}^{\infty}\frac{\vert\alpha\vert^{2n}}{n!~ h^2(n)}~~.
\end{equation}
Later we wish to construct a $q$-deformed Fock space of one-dimensional $q$-deformed oscillator algebra
\begin{equation}\label{qdefalgeb}
A_qA_q^\dagger-q^2A_q^\dagger A_q=1 \qquad q\leq 1
\end{equation}
in the basis $\vert n\rangle_q$ involving $q$-deformed integers $[n]_q$ as
\begin{eqnarray}
\vert n\rangle_q &:=& \frac{\left(A^\dagger\right)^n}{\sqrt{[n]_q!}}\vert 0\rangle_q, \qquad [n]_q!:= \displaystyle\prod_{k=1}^{n}[k]_q \\
A_q\vert 0\rangle_q &:=& 0 \qquad\quad ~_{q}\langle 0 \vert 0\rangle_q := 1~. \notag
\end{eqnarray}
Considering
\begin{equation}
[n]_q=\frac{1-q^{2n}}{1-q^2}~,
\end{equation}
it follows immediately that the operators $A_q^\dagger$ and $A_q$ act indeed as raising and lowering operators in the deformed space 
\begin{eqnarray}\label{qdefladder}
A_q^\dagger\vert n \rangle_q &=& \sqrt{[n+1]_q}~\vert n+1\rangle_q \\
A_q\vert n \rangle_q &=& \sqrt{[n]_q}~\vert n-1\rangle_q,\notag
\end{eqnarray}
which satisfy the deformed algebra (\ref{qdefalgeb}). The states $\vert n\rangle_q$ therefore form an orthonormal basis in $q$-deformed Hilbert space $\mathcal{H}_q$ spanned by the vectors $\vert \psi\rangle :=\sum_{n=0}^{\infty}c_n \vert n \rangle_q$ with $c_n\in\mathbb{C}$, such that $\langle \psi\vert\psi\rangle=\sum_{n=0}^{\infty}\vert c_n\vert^2<\infty$. 

For a concrete realisation of physical consequences of the deformed system, we can represent the deformed ladder operators in terms of the standard canonical observables 
\begin{eqnarray}\label{HermitianRep}
A_q &=& \frac{i}{\sqrt{1-q^2}}\left(e^{-i\hat{x}}-e^{-i\hat{x}/2}e^{2\tau \hat{p}}\right) \qquad \text{and} \\
A_q^\dagger &=& \frac{-i}{\sqrt{1-q^2}}\left(e^{i \hat{x}}-e^{2\tau \hat{p}}e^{i\hat{x}/2}\right), \notag
\end{eqnarray}
where $\hat{x}=\sqrt{m\omega/\hbar}$ and $\hat{p}=p/\sqrt{m \omega\hbar}$ are dimensionless observables with $x$, $p$ being canonical coordinates satisfying $[x,p]=i \hbar$ and the deformation parameter $q$ being parametrized to $q=e^\tau$. Clearly, $A_q^\dagger$ becomes Hermitian conjugate to $A_q$ for $q<1$ with respect to the given representation (\ref{HermitianRep}). Consequently, one can construct any physical Hamiltonian with the help of the ladder operators of the deformed system as per their requirement. For further interest in this context one might look at the previous works by the author \cite{Dey_Fring_Gouba_Castro,Dey_Fring_squeezed,Dey_Fring_Khantoul}, where the physical implications of the deformed noncommutative systems have been analysed thoroughly. We have also studied few other physical aspects of the deformed algebra (\ref{qdefalgeb}) in \cite{Dey_Fring_Gouba}, where it was shown that the noncommutative systems corresponding to the algebra (\ref{qdefalgeb}) led to the existence of minimal length as well as minimal momentum, which is a direct consequence of the string theory.

Having discussed the physical consequences of the algebra, we identify the ladder operators of the $q$-deformed noncommutative space (\ref{qdefladder}) with that of the generalised creation and annihilation operators (\ref{genladder}) with the choice
\begin{equation}
f(n)=\sqrt{\frac{[n+1]_q}{n+1}}~,
\end{equation} 
such that the $q$-deformed coherent states are computed from (\ref{noncoherent}) as
\begin{equation}\label{qdefcoherent}
\big\vert\alpha,f\big\rangle_q=\frac{1}{\mathcal{N}_q(\alpha,f)}\displaystyle\sum_{n=0}^{\infty}\frac{\alpha^n}{\sqrt{[n]_q!}}\vert n\rangle_q \qquad \alpha\in\mathbb{C},
\end{equation}
where the normalisation constant is represented in terms of the $q$-deformed exponential $E_q(\vert\alpha\vert^2)$
\begin{equation}
\mathcal{N}_q^2(\alpha,f)=\displaystyle\sum_{n=0}^{\infty}\frac{\vert\alpha\vert^{2n}}{[n]_q!}=E_q\big(\vert\alpha\vert^2\big)~~.
\end{equation}
%%%%%%%%%%%%%%%%%%%%%%%%%%%%%%%%%%%%%%%%%%%%%%%%%%%%%%%%%%%%%%%
\begin{subsection}{Classical-like properties of $q$-deformed coherent states}
In the previous section we have observed that the $q$-deformed coherent state (\ref{qdefcoherent}) can be constructed from the eigenvalue equation as well as by operating the generalised displacement operator on vacuum. In this section we will investigate some of their important properties, which indicate the classical-like behaviour of the noncommutative oscillator. Before we proceed, let us first define dimensionless quadrature operators
\begin{equation}\label{quadrature}
X=\frac{1}{2}\left(A_q+A_q^\dagger\right) \qquad \text{and} \qquad Y=\frac{1}{2i}\left(A_q-A_q^\dagger\right).
\end{equation} 
The expectation values of $A_q$ and $A_q^\dagger$ are computed as follows
\begin{equation}
_{q}\!\left\langle \alpha,f \right\vert A_q\left\vert \alpha,f \right\rangle_q=\alpha \qquad \text{and}\qquad _{q}\!\left\langle \alpha,f \right\vert A_q^\dagger\left\vert \alpha,f \right\rangle_q=\alpha^\ast,
\end{equation}
where $\alpha^\ast$ is the complex conjugate of $\alpha$. Therefore, we obtain
\begin{equation}
_{q}\!\left\langle \alpha,f \right\vert X\left\vert \alpha,f \right\rangle_q=\frac{(\alpha+\alpha^\ast)}{2} \qquad \text{and} \qquad _{q}\!\left\langle \alpha,f \right\vert Y\left\vert \alpha,f \right\rangle_q=\frac{(\alpha-\alpha^\ast)}{2i}~.
\end{equation}
To compute the expectation values of $X^2$ and $Y^2$, one requires
\begin{eqnarray}
_{q}\!\left\langle \alpha,f \right\vert A_qA_q\left\vert \alpha,f \right\rangle_q &=& \alpha^2 \\
_{q}\!\left\langle \alpha,f \right\vert A_q^\dagger A_q^\dagger\left\vert \alpha,f \right\rangle_q &=& \left(\alpha^\ast\right)^2 \\
_{q}\!\left\langle \alpha,f \right\vert A_q^\dagger A_q\left\vert \alpha,f \right\rangle_q &=& \left\vert\alpha\right\vert^2 \\
_{q}\!\left\langle \alpha,f \right\vert A_qA_q^\dagger\left\vert \alpha,f \right\rangle_q &=& 1+q^2\left\vert\alpha\right\vert^2.
\end{eqnarray}
With $X^2=\frac{1}{4}(A_qA_q+A_qA_q^\dagger+A_q^\dagger A_q+A_q^\dagger A_q^\dagger),~Y^2=-\frac{1}{4}(A_qA_q-A_qA_q^\dagger-A_q^\dagger A_q+A_q^\dagger A_q^\dagger)$ and $[X,Y]=\frac{i}{2}\big[1+(q^2-1) A_q^\dagger A_q\big]$, we assemble these to obtain 
\begin{equation}
(\Delta X)^2\Big\vert_{\vert\alpha,f\rangle_q}=(\Delta Y)^2\Big\vert_{\vert\alpha,f\rangle_q}=\frac{1}{2}\Big\vert_{q}\!\big\langle \alpha,f \big\vert [X,Y]\big\vert \alpha,f \big\rangle_q\Big\vert=\frac{1}{4}\Big\{1+\left(q^2-1\right)\vert\alpha\vert^2\Big\},
\end{equation}
where the variance of $X$ is computed as $\langle(\Delta X)^2\rangle=_{q}\!\left\langle \alpha,f \right\vert X^2\left\vert \alpha,f \right\rangle_q-_{q}\!\left\langle \alpha,f \right\vert X\left\vert \alpha,f \right\rangle_q^2$ and similarly for the other quadrature $Y$. The generalised uncertainty relation
\begin{eqnarray}\label{GUR}
\Delta X~\Delta Y\Big\vert_{\vert\alpha,f\rangle_q} &\geq & \frac{1}{2}\Big\vert_{q}\!\big\langle \alpha,f \big\vert [X,Y]\big\vert \alpha,f \big\rangle_q\Big\vert
\end{eqnarray}
is therefore saturated in this case. This is the reason why coherent states are called minimum uncertainty states. Moreover, the uncertainties of the two quadratures $X$ and $Y$ are identical to each other. Note that, this is essentially true for the vacuum state as well, which means coherent states produce equal amounts of optical noise as vacuum states. For further details in this context, see \cite{Gerry_Knight_Book}. In principle, coherent states can be constructed in the reverse path; for instance, one could define coherent states as states that minimize the uncertainty relation for the two orthogonal field quadratures with equal uncertainties in each quadrature and construct an explicit expression of coherent state out of it \cite{Gerry_Knight_Book}. However, it is important to note that all generalised coherent states do not minimize the uncertainty relation, as we have noticed in one of our recent works \cite{Dey_Fring_Gouba_Castro}, where we considered Gazeau-Klauder coherent states \cite{Klauder1995,Gazeau_Klauder} in our analysis. We found that the uncertainty relation is saturated in that case only at a particular value of time. In this sense, nonlinear coherent states might be more useful than Gazeau-Klauder coherent states for this purpose.  
\end{subsection}
%%%%%%%%%%%%%%%%%%%%%%%%%
\begin{subsubsection}{Sub-Poissonian behaviour}
To examine the sub-Poissonian behaviour of the $q$-deformed coherent states, let us first compute the photon number average and its dispersion
\begin{eqnarray}
_q\langle n\rangle_q &=& _{q}\!\left\langle \alpha,f \right\vert A_q^\dagger A_q\left\vert \alpha,f \right\rangle_q = \left\vert\alpha\right\vert^2 \label{na}\\
(\Delta n)^2\Big\vert_{\vert\alpha,f\rangle_q} &=& _{q}\!\left\langle \alpha,f \right\vert A_q^\dagger A_qA_q^\dagger A_q\left\vert \alpha,f \right\rangle_q - _{q}\!\left\langle \alpha,f \right\vert A_q^\dagger A_q\left\vert \alpha,f \right\rangle_q= \left\vert\alpha\right\vert^2 +q^2\left\vert\alpha\right\vert^4. \label{nsa}
\end{eqnarray} 
For a measurement of the number of photons, the probability of detecting $n$ photons is
\begin{eqnarray}\label{photoncoherent}
\mathcal{P}_{n,q} &=& \big\vert  _{q}\!\left\langle n\vert \alpha,f \right\rangle_q\big\vert^2~=~\Bigg\vert\frac{\alpha^n}{\mathcal{N}_q(\alpha,f)\sqrt{[n]_q!}}\Bigg\vert^2,
\end{eqnarray}
whose behaviour is clearly sub-Poissonian as shown in figure \ref{fig4}. However, for a concrete realisation of the behaviour one can compute the Mandel parameter $Q$ \cite{Mandel}
\begin{eqnarray}\label{Mandel}
Q &=& \frac{(\Delta n)^2}{\langle n\rangle}-1,
\end{eqnarray}
which identifies the nature of the photon probability distribution function (\ref{photoncoherent}). $Q=0$ corresponds to the case of Poissonian distribution, whereas $Q>0$ and $Q<0$ imply to the super-Poissonian and sub-Poissonian cases respectively. For the case at hand, we compute the Mandel parameter (\ref{Mandel}) with the help of equations (\ref{na}) and (\ref{nsa}) to 
\begin{eqnarray}\label{qdefMandel}
Q_q=(q^2-1)\left\vert\alpha\right\vert^2 \qquad q\leq 1~.
\end{eqnarray}
Note that in the case when $q=1$, which corresponds to the Glauber coherent states, the Mandel parameter turns out to be zero. Which means, for ordinary coherent states the photon distribution is always Poissonian and therefore number squeezing is absent in that case \cite{Gerry_Knight_Book}. However, if one considers the $q$-deformed case (\ref{qdefMandel}) and restricts $-1<q<1$ further, the photon distribution remains sub-Poissonian. Thus this is quite obvious that the parameter $q$ supplies one degree of freedom in the scenario and therefore the q-deformed coherent states is one step further in quality than the ordinary Glauber states as because of the number squeezing.
\end{subsubsection}
\end{section} 

%%%%%%%%%%%%%%%%%%%%%%%%%%%%%%%%%%%%%%%%%%%%%%%%%%%%%%%%%%%%%%%%%%%%%%%%%%%%%%%%%%%%%%%%%%%%%%%%%%%%
%  Section 3
%%%%%%%%%%%%%%%%%%%%%%%%%%%%%%%%%%%%%%%%%%%%%%%%%%%%%%%%%%%%%%%%%%%%%%%%%%%%%%%%%%%%%%%%%%%%%%%%%%%%

\begin{section}{$q$-deformed cat states}\label{sec3}
In the previous section we have looked at different classical-like properties that coherent states inherit and noticed that the wave packet behaves indeed as a classical soliton-like particle. Let us now consider the superposition of two coherent states which are of equal amplitude but different phases
\begin{eqnarray}
\big\vert\alpha,f,\pm\big\rangle_q &=& \frac{1}{\mathcal{N}_q(\alpha,f,\pm)}\Big(\vert\alpha,f\rangle_q\pm\vert-\alpha,f\rangle_q\Big),
\end{eqnarray} 
with the normalisation constant
\begin{eqnarray}
\mathcal{N}_q^2(\alpha,f,\pm) &=& 2\pm\frac{2}{\mathcal{N}_q^2(\alpha,f)}\displaystyle\sum_{n=0}^{\infty}\frac{(-1)^n\vert\alpha\vert^{2n}}{[n]_q!}~=~2\Big\{1\pm E_q(-2\vert\alpha\vert^2)\Big\},
\end{eqnarray}
which are familiar as Schr{\"o}dinger cat states, however $q$-deformed in this case. Because even and odd photons are combined in the states $\vert\alpha,f,+\rangle_q$ and $\vert\alpha,f,-\rangle_q$  respectively, sometimes the corresponding states are also known as even and odd coherent states \cite{Xia_Guo,Filho_Vogel1,Mancini,Sivakumar1}.
%%%%%%%%%%%%%%%%%%%%%%%%%%%%%%%%%%%%%%%%%%%%%%%%%%%%%%%%
\begin{subsection}{Nonclassical properties of $q$-deformed cat states}
Before we start let us make our terminology clear to the readers. One might demand the actual meaning of nonclassicality in this context. Eventually, one could ask aren't all states of light including coherent states quantum mechanical? The answer is yes but it turns out that some states are more quantum mechanical than others. Let us speak more clearly. Glauber and Sudarshan introduced their $P$-function \cite{Glauber2,Sudarshan2} as a probability distribution where one could expect $P\geq 0$, which is indeed true for coherent states. In fact, for coherent states the $P$-function is a delta function. However, for some quantum mechanical states they can be negative or highly singular than a delta function. In that sense those states, which correspond to the negative $P$-function in some region of space or are more singular than a delta function are often called nonclassical states \cite{Gerry_Knight_Book,Gerry}. Squeezed states are highly nonclassical in this sense \cite{Loudon_Knight}. 
\begin{subsubsection}{Quadrature Squeezing}
We consider the two quadrature operators which were defined in (\ref{quadrature}) to compute the dispersion of the corresponding quadratures. The expectation values of the creation and annihilation operators turn out to be zero in this case, so that one obtains
\begin{eqnarray}
_{q}\!\left\langle \alpha,f,\pm \right\vert X\left\vert \alpha,f,\pm \right\rangle_q &=& _{q}\!\left\langle \alpha,f,\pm \right\vert Y\left\vert \alpha,f,\pm \right\rangle_q~=~0~~.
\end{eqnarray}
However, the expectation values of combinations of $A_q$s and $A_q^\dagger$s do exist
\begin{eqnarray}
_{q}\!\left\langle \alpha,f,\pm \right\vert A_qA_q\left\vert \alpha,f,\pm \right\rangle_q &=& \alpha^2 \\
_{q}\!\left\langle \alpha,f,\pm \right\vert A_q^\dagger A_q^\dagger\left\vert \alpha,f,\pm \right\rangle_q &=& (\alpha^\ast)^2 \\
_{q}\!\left\langle \alpha,f,\pm \right\vert A_q^\dagger A_q\left\vert \alpha,f,\pm \right\rangle_q &=& \vert\alpha\vert^2 F_{q,\pm} \label{qdefnumber}\\
_{q}\!\left\langle \alpha,f,\pm \right\vert A_q A_q^\dagger\left\vert \alpha,f,\pm \right\rangle_q &=& 1+q^2\vert\alpha\vert^2 F_{q,\pm} \\
_{q}\!\left\langle \alpha,f,\pm \right\vert A_q^\dagger A_q A_q^\dagger A_q\left\vert \alpha,f,\pm \right\rangle_q &=& \vert\alpha\vert^2+q^2\vert\alpha\vert^4 F_{q,\pm}~,
\end{eqnarray}
where we introduce 
\begin{eqnarray}
F_{q,\pm} &:=& \frac{1\mp E_q(-2\vert\alpha\vert^2)}{1\pm E_q(-2\vert\alpha\vert^2)}~,
\end{eqnarray}
which when assembled, one obtains
\begin{eqnarray}
_{q}\!\left\langle \alpha,f,\pm \right\vert X^2\left\vert \alpha,f,\pm \right\rangle_q &=& \frac{1}{4}\Big\{1+\alpha^2+(\alpha^\ast)^2+(q^2+1)\vert\alpha\vert^2 F_{q,\pm}\Big\} \\
_{q}\!\left\langle \alpha,f,\pm \right\vert Y^2\left\vert \alpha,f,\pm \right\rangle_q &=& \frac{1}{4}\Big\{1-\alpha^2-(\alpha^\ast)^2+(q^2+1)\vert\alpha\vert^2F_{q,\pm}\Big\}~.
\end{eqnarray}
The square of the uncertainties are therefore
\begin{eqnarray}
(\Delta X)^2\Big\vert_{\left\vert \alpha,f,\pm \right\rangle_q} &=& G_q+\frac{1}{4}\Big\{\alpha^2+(\alpha^\ast)^2+2\vert\alpha\vert^2F_{q,\pm}\Big\} \label{DXS}\\
(\Delta Y)^2\Big\vert_{\left\vert \alpha,f,\pm \right\rangle_q} &=& G_q-\frac{1}{4}\Big\{\alpha^2+(\alpha^\ast)^2-2\vert\alpha\vert^2F_{q,\pm}\Big\}, \label{DYS}
\end{eqnarray}
where
\begin{eqnarray}\label{Gq}
G_q &=& \frac{1}{4}\Big\{1+(q^2-1)\vert\alpha\vert^2F_{q,\pm}\Big\}~.
\end{eqnarray}
The square of left-hand side of the generalised uncertainty relation (\ref{GUR}) is obtained from (\ref{DXS}) and (\ref{DYS})
\begin{eqnarray}\label{GURLHS}
(\Delta X)^2(\Delta Y)^2\Big\vert_{\left\vert \alpha,f,\pm \right\rangle_q} &=& G_q^2+G_q \vert\alpha\vert^2 F_{q,\pm}-\frac{\big\{\alpha^2+(\alpha^\ast)^2\big\}^2}{16}+\frac{\vert\alpha\vert^4}{4}F_{q,\pm}^2~,
\end{eqnarray}
whereas the square of the right-hand side is acquired from (\ref{qdefnumber}) 
\begin{eqnarray}\label{GURRHS}
\frac{1}{4}\Big\vert_{q}\!\big\langle \alpha,f,\pm \big\vert [X,Y]\big\vert \alpha,f,\pm \big\rangle_q\Big\vert^2 &=& \frac{1}{16}\Big\{1+(q^2-1)\vert\alpha\vert^2F_{q,\pm}\Big\}^2~=~G_q^2~.
\end{eqnarray}
Therefore the uncertainty relation is always true if the following condition holds
\begin{eqnarray}\label{GURcondition}
G_q \vert\alpha\vert^2 F_{q,\pm}+\frac{\vert\alpha\vert^4}{4}F_{q,\pm}^2 &\geq& \frac{\big\{\alpha^2+(\alpha^\ast)^2\big\}^2}{16}~.
\end{eqnarray}
One can easily check that the condition (\ref{GURcondition}) is consistent for any values of $\alpha$ and $q$ for odd cat states, whereas for the even case not all values of $q$ and $\alpha$ are allowed. In that case one needs to
\begin{figure}[H]
\centering   \includegraphics[width=8.2cm,height=6.0cm]{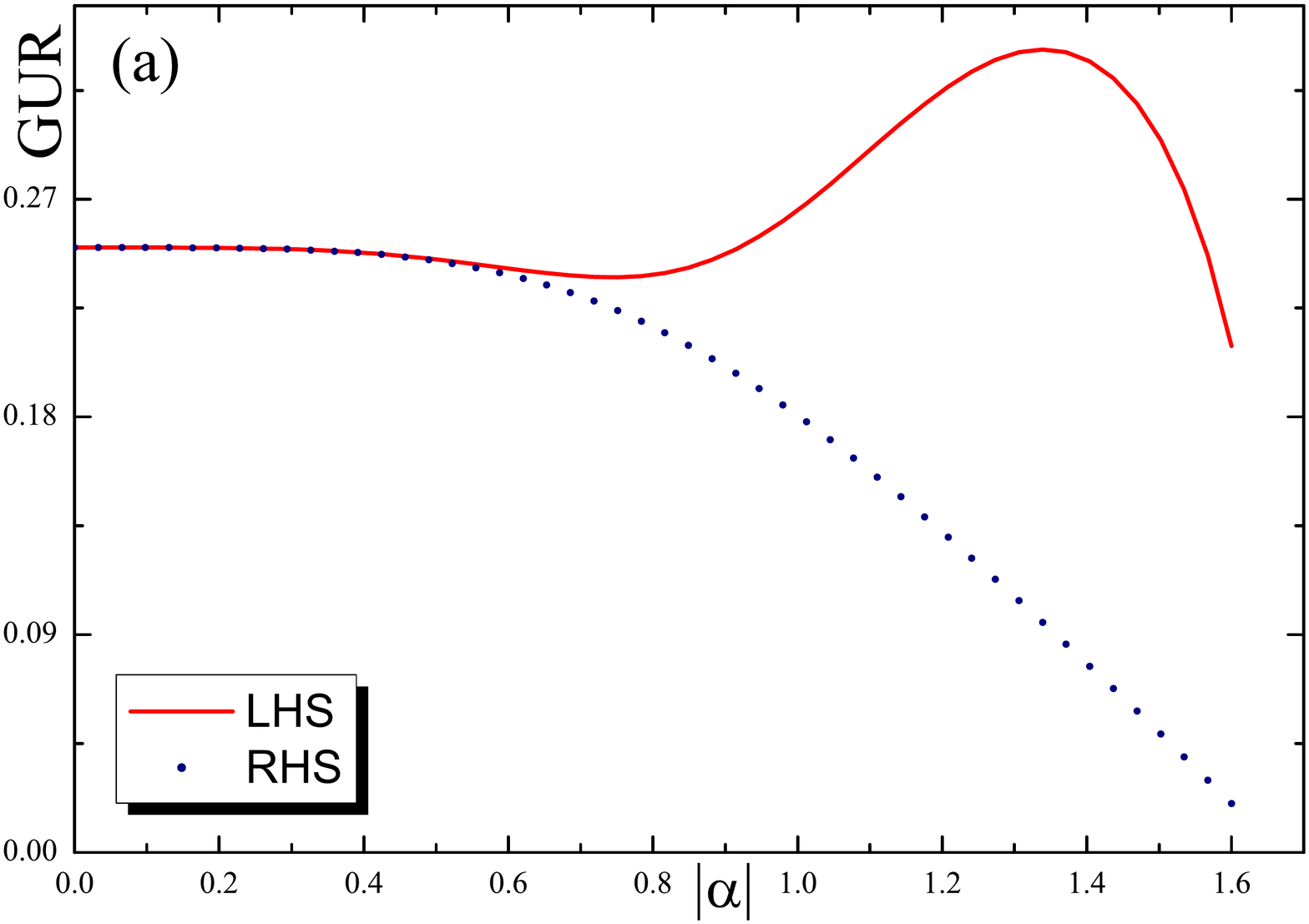}
\includegraphics[width=8.2cm,height=6.0cm]{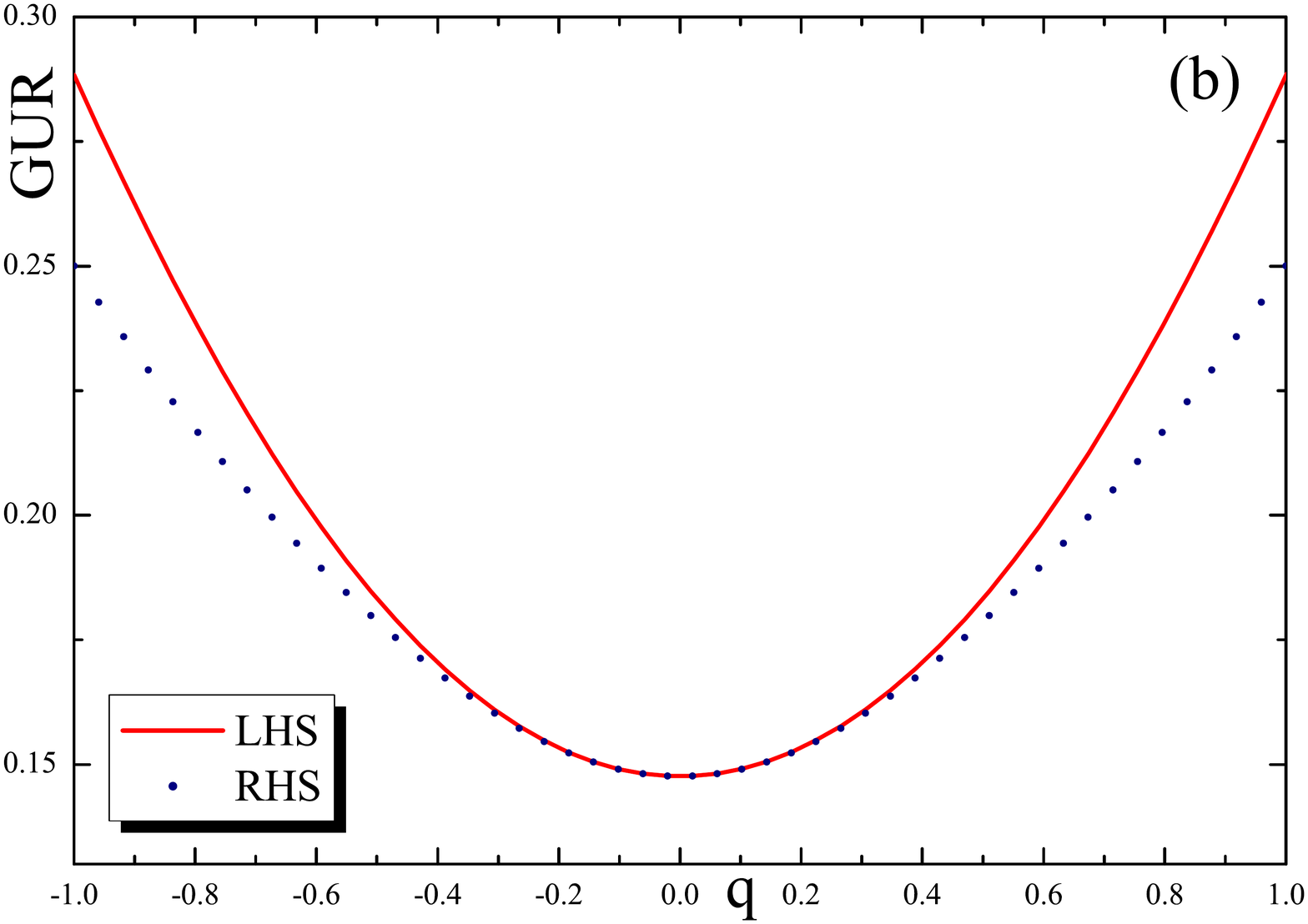}
\caption{\small{Left-hand side (\ref{GURLHS}) and right-hand side (\ref{GURRHS}) of generalised uncertainty relation for the even cat state (a) for $q=0.8$ as a function of $\vert\alpha\vert$  (b) for $\vert\alpha\vert=0.8$ as a function of $q$.}}
\label{fig1}
\end{figure}
\begin{figure}
\centering   \includegraphics[width=8.2cm,height=6.0cm]{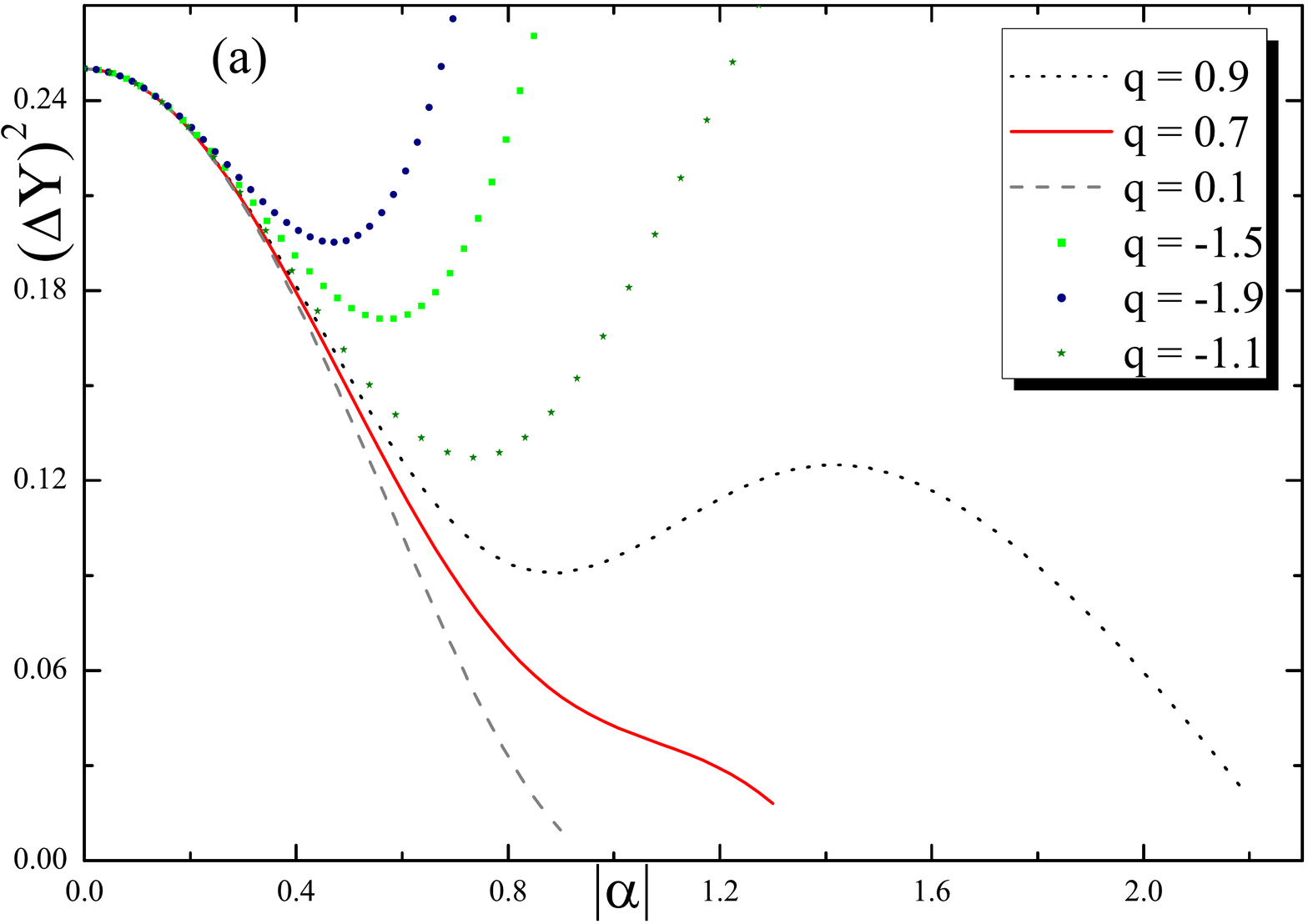}
\includegraphics[width=8.2cm,height=6.0cm]{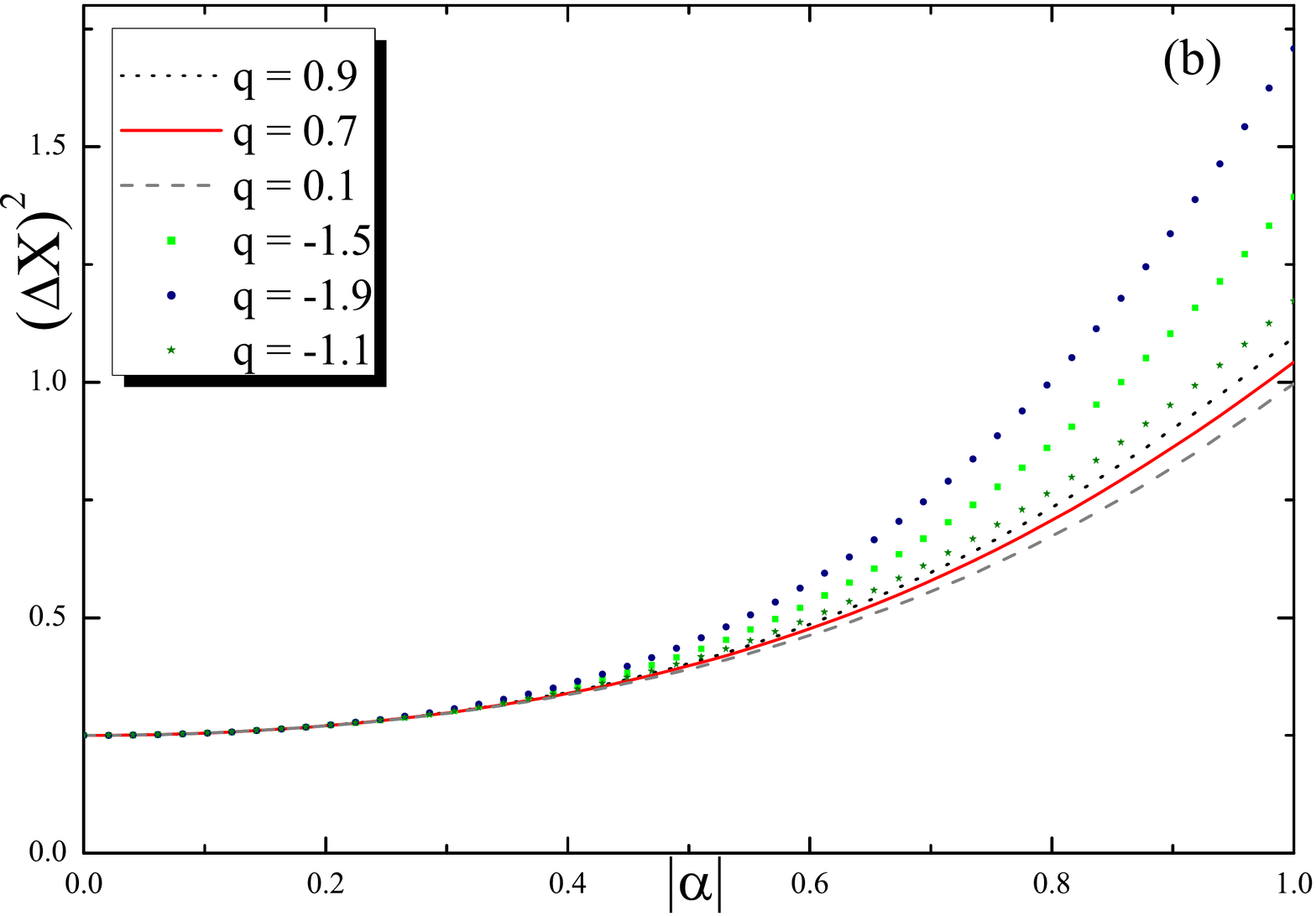}
\caption{\small{Uncertainty in (a) $Y$ quadrature (b) $X$ quadrature as a function of $\vert\alpha\vert$ for even cat states.}}
\label{fig2}
\end{figure}
\noindent restrict the values of $q$ and/or $\alpha$ in such a way that the condition (\ref{GURcondition}) satisfies. We have compared left-hand side (\ref{GURLHS}) with right-hand side (\ref{GURRHS}) of the uncertainty relation inside the allowed range of $\alpha$ and $q$ which are shown in figure \ref{fig1}. Notice that unlike the coherent states, uncertainties in two quadratures are not equal to each other; rather, in even cat states the quadrature $Y$ (\ref{DYS}) is squeezed below the right-hand side (\ref{GURRHS}) of the uncertainty relation, whereas the quadrature $X$ (\ref{DXS}) is expanded correspondingly such that the uncertainty relation holds, as shown in figure \ref{fig2}. From equations (\ref{DXS}-\ref{Gq}) we can compute a condition for which the squeezing in $Y$ quadrature exists for even cat states
\begin{eqnarray}
\frac{E_q(-2\vert\alpha\vert^2)Re[\alpha]^2-Im[\alpha]^2}{4\{1+E_q(-2\vert\alpha\vert^2)\}}\Big\{\alpha^2+(\alpha^\ast)^2-2(1+q^2\vert\alpha\vert^2F_{q,+})\Big\} &<& 0~.
\end{eqnarray}
For odd cat states no such squeezing exists. Squeezing in quadratures means producing less amount of noise than in coherent states \cite{Loudon_Knight}. More interesting effects can be observed from figure \ref{fig3}, where the squeezing in $Y$ quadrature is plotted as a function of the noncommutative parameter $q$ for a fixed value of $\vert\alpha\vert$. It is obvious that the presence of an extra parameter provides an additional freedom to squeeze the quadrature beyond the usual case ($q=1$). Moreover, one can adjust the values of $\alpha$ and $q$ in such a way that one can obtain minimum uncertainty states. The minimum uncertainty states with squeezing in one quadrature are sometimes called "ideal squeezed states" or "intelligent states" \cite{Gerry_Knight_Book}, which are achievable from noncommutative space, for instance for $q=0.9$ and $\vert\alpha\vert=2.2648$.  
\begin{figure}
\centering   \includegraphics[width=8.2cm,height=6.0cm]{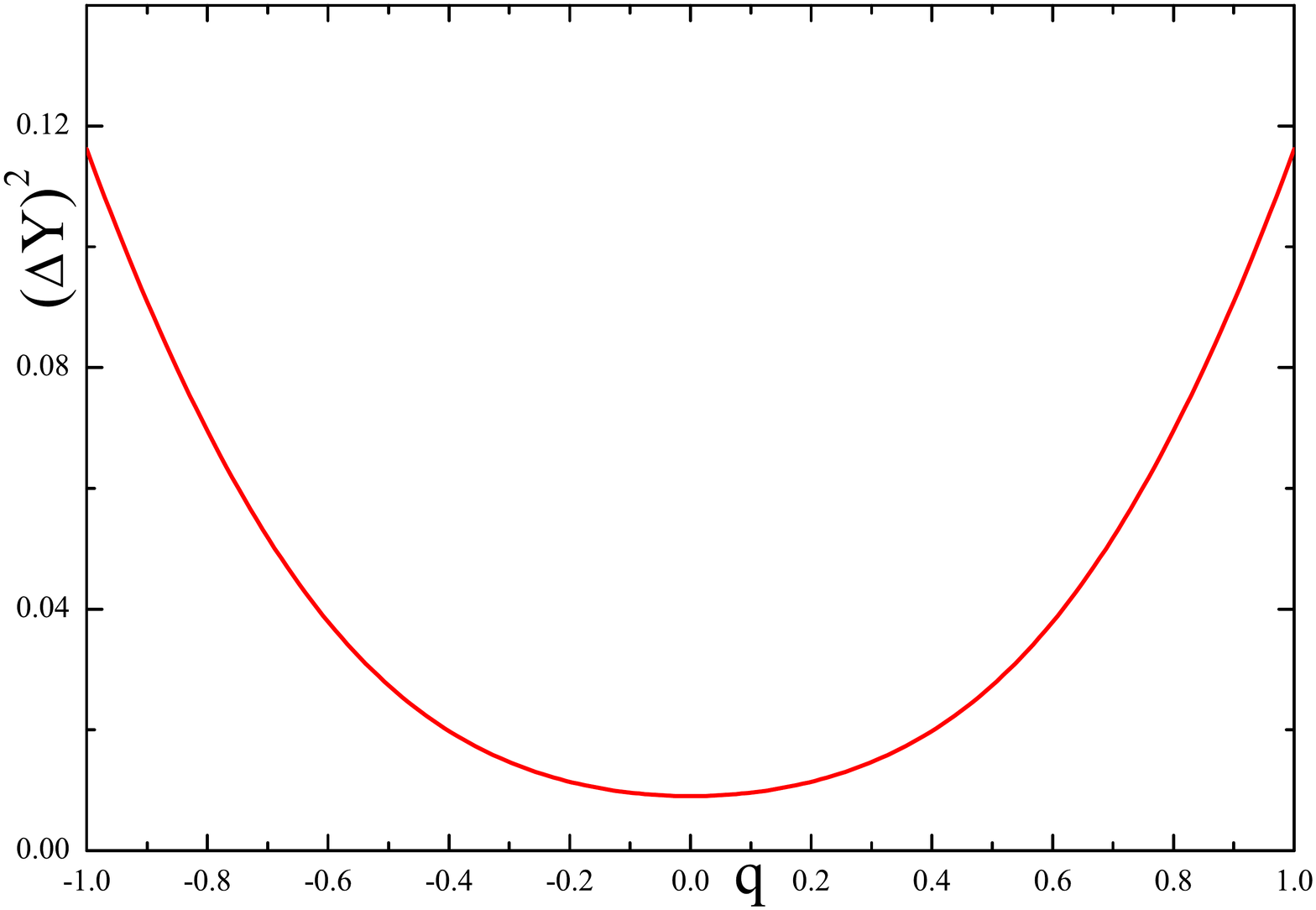}
\caption{\small{Uncertainty in $Y$ quadrature as a function of $q$ for even cat states for $\vert\alpha\vert=0.9$.}}
\label{fig3}
\end{figure}
\end{subsubsection}
%%%%%%%%%%%%%%%%%%%%%%%%%%%%%%%%%
\begin{subsubsection}{Photon distribution and number squeezing}
Let us now compute the photon number dispersion
\begin{eqnarray}
(\Delta n)^2\Big\vert_{\left\vert \alpha,f,\pm \right\rangle_q} &=& \vert\alpha\vert^2\Big\{1+\vert\alpha\vert^2F_{q,\pm}\big(q^2-F_{q,\pm}\big)\Big\}
\end{eqnarray}
and the photon distribution function for this case
\begin{eqnarray}\label{photoncat}
\mathcal{P}_{n,q,\pm} &=& \big\vert  _{q}\!\left\langle n\vert \alpha,f,\pm \right\rangle_q\big\vert^2~=~\Bigg\vert\frac{1}{\mathcal{N}_q(\alpha,f,\pm)\mathcal{N}_q(\alpha,f)}\Bigg(\frac{\alpha^n}{\sqrt{[n]_q!}}+\frac{(-1)^n\alpha^n}{\sqrt{[n]_q!}}\Bigg)\Bigg\vert^2.
\end{eqnarray}
\begin{figure}[h]
\centering   \includegraphics[width=8.2cm,height=6.0cm]{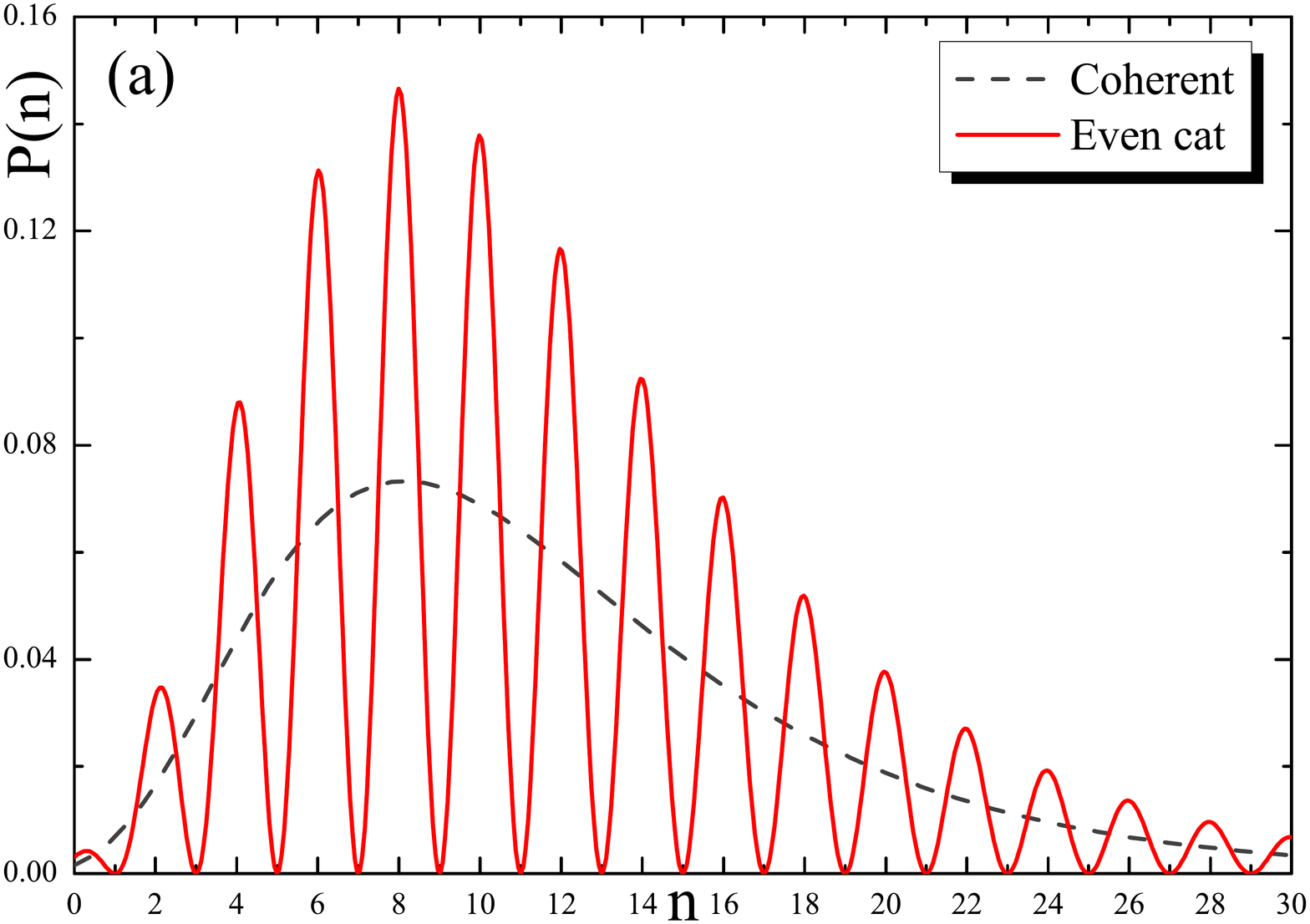}
\includegraphics[width=8.2cm,height=6.0cm]{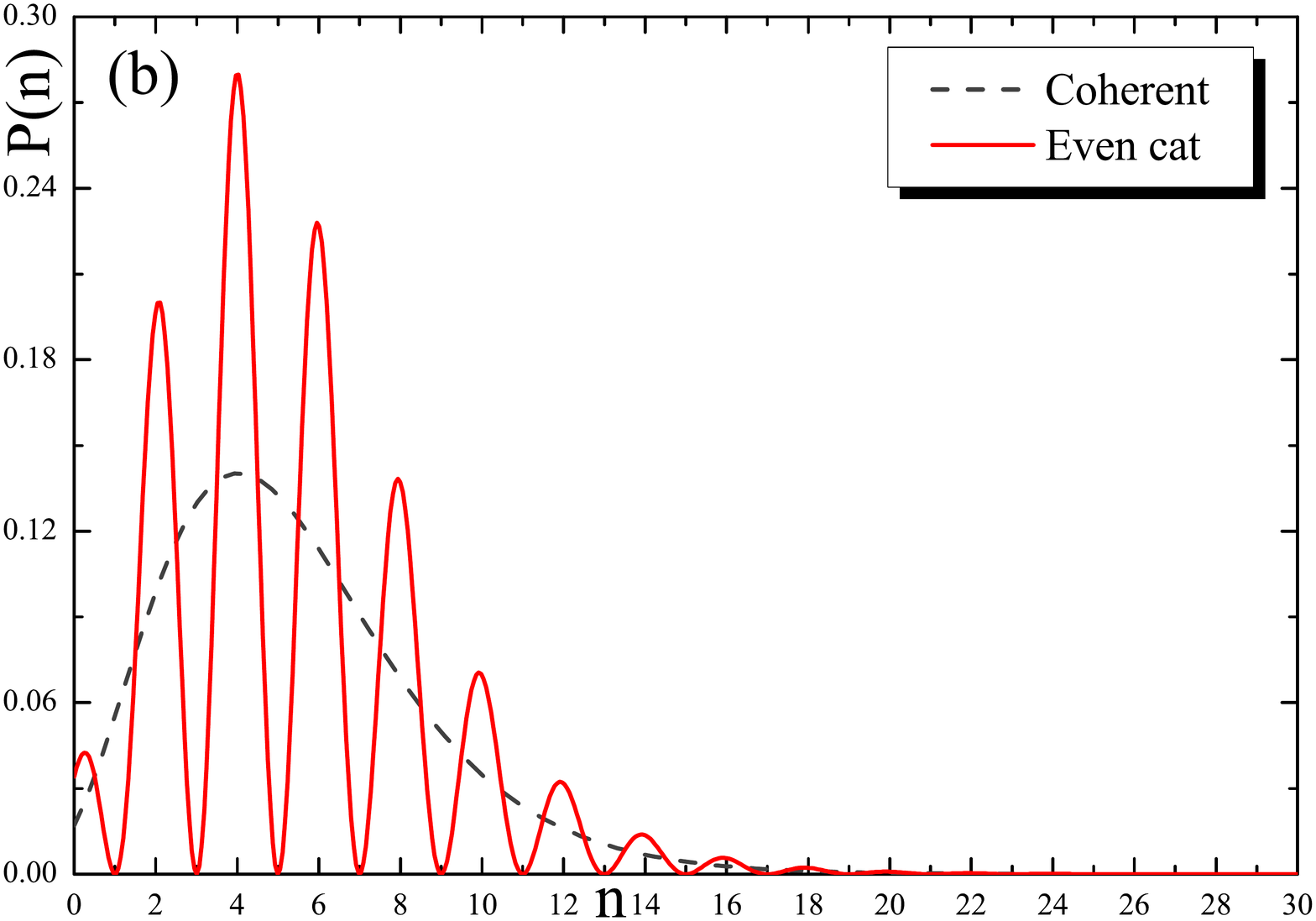}
\caption{\small{Photon distribution functions of coherent state (grey dashed) and even cat state (red solid) as a function of the number of photons (n) for $q=0.9$ (a) $\vert\alpha\vert=2.1$ (b) $\vert\alpha\vert=1.8$.}}
\label{fig4}
\end{figure}
It is evident that the probability of finding odd (even) photons in the even (odd) cat states vanishes and thus the photon distribution function (\ref{photoncat}) becomes highly oscillatory, which is a strong evidence of nonclassicality \cite{Gerry_Knight_Book}. In figure \ref{fig4} we have compared the number distribution function of even cat states (\ref{photoncat}) with that of the coherent states (\ref{photoncoherent}), which clearly shows the classical-like nature of coherent states in contrast to the nonclassical behaviour of cat states. Let us now look at the Mandel parameter (\ref{Mandel}) for this case
\begin{eqnarray}\label{Mandelcat}
Q_{q,\pm} &=& \frac{1}{F_{q,\pm}}-1+\big(q^2-F_{q,\pm}\big)\vert\alpha\vert^2.
\end{eqnarray}
For odd cat states this is always negative (sub-Poissonian) irrespective of the values of $\alpha$ and $q$ which are depicted in panel (b) of figure \ref{fig5}. This suggests that because of photon antibunching, the photon number distribution is squeezed in this case. The advantage of choosing noncommutative space is clearly visible here; for instance, the photon distribution function becomes more and more negative with the increase of the value of $\vert\alpha\vert$, which does not happen for the ordinary case as depicted by the scattered (dark grey) line in panel (b) of figure \ref{fig5}. However, the quadratures are not squeezed for the odd cat states. More interesting are the even states, where the $Y$ quadrature is squeezed as well as the photon number distribution albeit for some values of $\alpha$ and $q$ as shown in panel (a) of figure \ref{fig5}. The photons are therefore antibunched for the even cat states in contrast to the ordinary algebra where the antibunching effect happens only for the odd cat states \cite{Hillery,Xia_Guo}. In the ordinary quantum mechanical case corresponding to $q=1$, we compute the Mandel parameter
\begin{eqnarray}\label{QOrdinary}
Q_\pm &=& \frac{2}{1-e^{4\vert\alpha\vert^2}}\Big\{\vert\alpha\vert^2-1\mp e^{2\vert\alpha\vert^2}\big(1+\vert\alpha\vert^2\big)\Big\},
\end{eqnarray}
\begin{figure}[H]
\centering   \includegraphics[width=8.2cm,height=6.0cm]{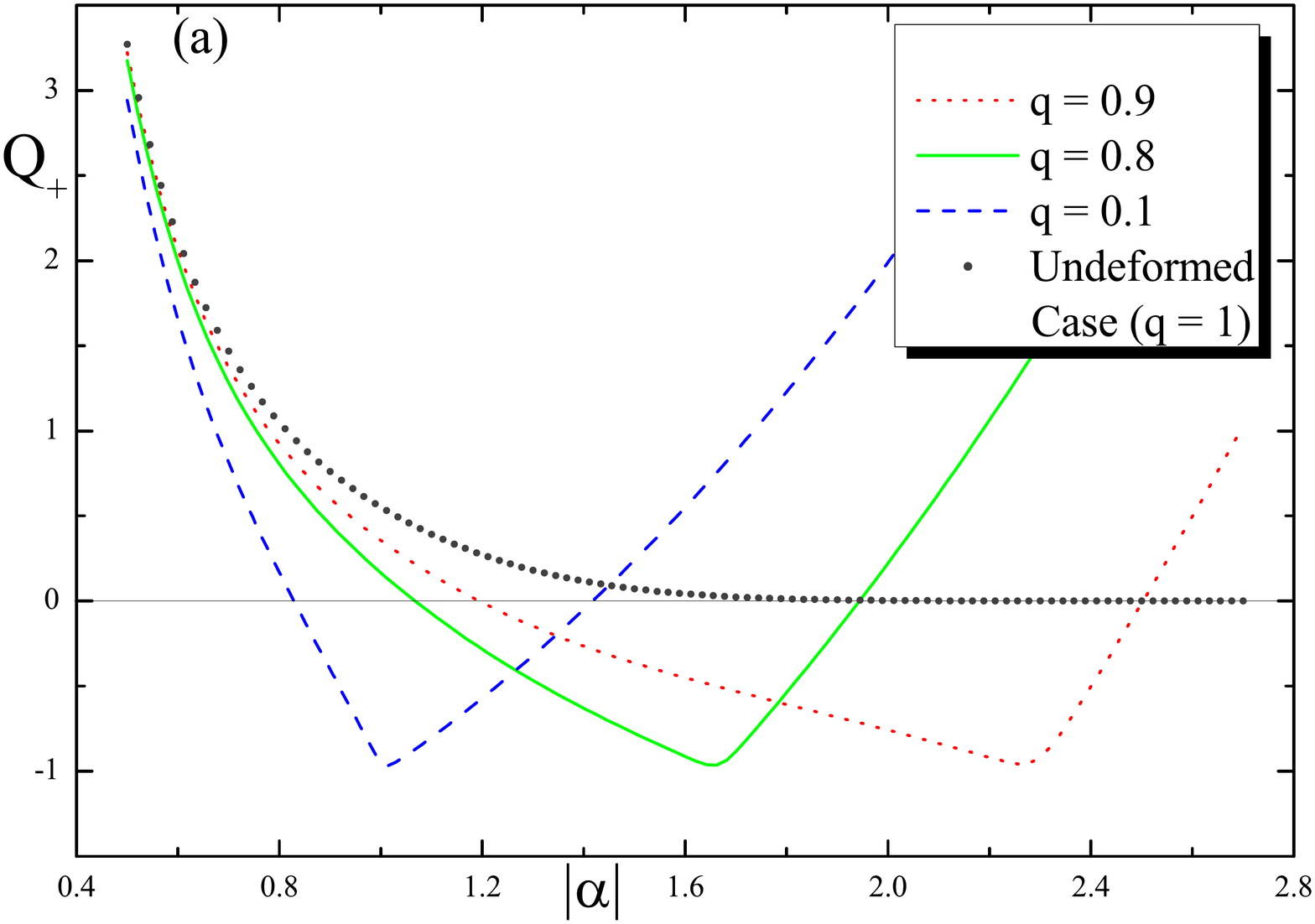}
\includegraphics[width=8.2cm,height=6.0cm]{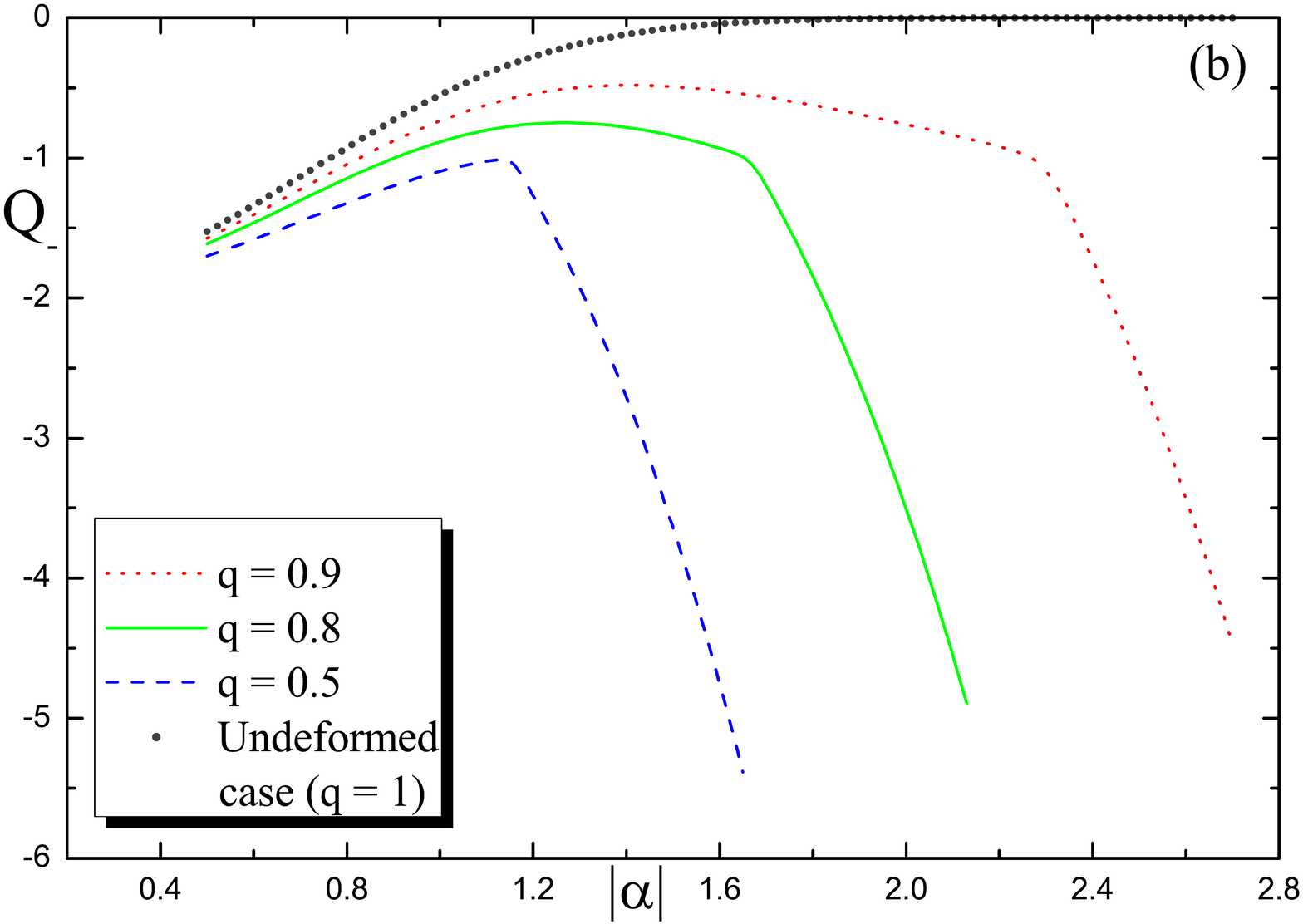}
\caption{\small{Mandel parameter for $q$-deformed cat states (\ref{Mandelcat}) (blue dashed, green solid and red dotted) and ordinary cat states (\ref{QOrdinary}) (dark grey scattered) as a function of $\vert\alpha\vert$. (a) Even cat states. (b) Odd cat states.}}
\label{fig5}
\end{figure}
\noindent which is however always positive for the even cat states as shown in panel (a) of figure \ref{fig5}. In the usual cat states the photon number distribution are therefore broader than the coherent states. On the other hand, it is possible to obtain a squeezed photon distribution than a coherent state along with the squeezing in $Y$ quadrature for the even case of $q$-deformed cat states.
\end{subsubsection}
\end{subsection}  
\end{section}

%%%%%%%%%%%%%%%%%%%%%%%%%%%%%%%%%%%%%%%%%%%%%%%%%%%%%%%%%%%%%%%%%%%%%%%%%%%%%%%%%%%%%%%%%%%%%%%%%%%%
%  Section 4
%%%%%%%%%%%%%%%%%%%%%%%%%%%%%%%%%%%%%%%%%%%%%%%%%%%%%%%%%%%%%%%%%%%%%%%%%%%%%%%%%%%%%%%%%%%%%%%%%%%%

\begin{section}{Conclusions}\label{sec4}
It is amusing that the superposition of two coherent states shows many nonclassical properties. We have analysed some of them here in $q$-deformed noncommutative space, which are always more complicated and challenging from the computational point of view. However, one always has the possibility of obtaining more degrees of freedom in such spaces for instance via the parameter $q$ in our case. We have shown that the parameter $q$ can be utilised to obtain minimum uncertainty quadrature squeezed states, i.e. the intelligent states which is difficult to achieve with the ordinary systems. Moreover, we have indicated that it is possible to acquire both quadrature squeezing and number squeezing at the same time for the even cat states in the $q$-deformed space which has never happened in the usual space, where one has either quadrature or number squeezing but not both of them together. It is also observed that the nonclassicality of the system can be improved considerably in noncommutative space with the insertion of the parameter $q$. For instance, the $Y$ quadrature can be squeezed further in the even case, while in the odd case the photon distribution can be made more narrow than in the ordinary quantum mechanical cat states.

Properties of $q$-deformed coherent states have been studied alongside. Nonlinear coherent states have been realized to be minimum uncertainty states in the $q$-deformed space which are not always true for instance for Gazeau-Klauder coherent states in the same space \cite{Dey_Fring_Gouba_Castro}. The wave packets in ordinary coherent states are always Poissonian \cite{Gerry_Knight_Book}, whereas in noncommutative systems they can be adjusted to make sub-Poissonian, which demonstrates that the $q$-deformed coherent states create less amount of optical noise than the Glauber coherent states, which was also realized partly in our previous investigations \cite{Dey_Fring_squeezed,Dey_Fring_Gouba_Castro}.

There are many directions left for further investigation. First, one can choose many other noncommutative systems to verify our observations. Second, it will be very exciting to study few other nonclassical properties to confirm our findings; for instance, one could compute the Wigner function \cite{Wigner} for the system we considered here, which must have negative peaks in some regions of space. 
\begin{subsubsection}*{Acknowledgements:}
S. D. is supported by the Post Doctoral Fellowship jointly funded by the Laboratory of Mathematical Physics of the Centre de Recherches Math{\'e}matiques (CRM) and by Prof. Twareque Ali, Prof. Marco Bertola and Prof. V{\'e}ronique Hussin.
\end{subsubsection}
\end{section}

%%%%%%%%%%%%%%%%%%%%%%%%%%%%%%%%%%%%%%%%%%%%%%%%%%%%%%%%%%%%%%%%%%%%%%%%%%%%%%%%%%%%%%%%%%%%%%%%%%%%

%\bibliographystyle{unsrt}
%\bibliography{nccat.bib}

\end{document}